\documentclass{article}
\usepackage{ijcai17}

\usepackage{times}
\usepackage{helvet}
\usepackage{courier}
\usepackage{graphicx}
\usepackage{amssymb}
\usepackage{amsmath}
\usepackage{subfigure}
\usepackage{algorithm}
\usepackage[noend]{algorithmic}
\usepackage{url}
\usepackage{pifont}
\usepackage{color}

\newcommand{\bphi}{\boldsymbol{\phi}}
\newcommand{\bpsi}{\boldsymbol{\psi}}

\newcommand{\stitle}[1]{\vspace{2mm} \noindent {\bf #1}}
\newcommand{\eat}[1]{}
\newcommand{\eg}{{\it e.g.}}

\newcommand{\ie}{{\it i.e.}}

\newcommand{\wrt}{{\it w.r.t. }}

\newtheorem{definition}{Definition}

\title{From Node Embedding To Community Embedding\thanks{This is a technical report for paper [Cavallari \emph{et al.}, 2017].}}
\author{
Vincent W. Zheng$^{1}$, Sandro Cavallari$^{2}$, Hongyun Cai$^{1}$, Kevin Chen-Chuan Chang$^{3}$, Erik Cambria$^{2}$\\
$^{1}$ Advanced Digital Sciences Center, Singapore; \\
$^{2}$ Nanyang Technological University, Singapore;  \\
$^{3}$ University of Illinois at Urbana-Champaign, USA \\
\{vincent.zheng, hongyun.c\}@adsc.com.sg, sandro001@e.ntu.edu.sg, \\ kcchang@illinois.edu, cambria@ntu.edu.sg
}

\begin{document}
\maketitle
\begin{abstract}
In this paper, we introduce a new setting for graph embedding, which considers embedding communities instead of individual nodes. 
Community embedding is useful as a natural community representation for applications, and it provides an exciting opportunity to improve community detection. 
Specifically, we see the interaction between community embedding and detection as a closed loop, through node embedding. 
On the one hand, we rely on node embedding to generate good communities and thus meaningful community embedding. On the other hand, we apply community embedding to improve node embedding through a novel community-aware higher-order proximity. 
This closed loop enables us to improve community embedding, community detection and node embedding at the same time. 
Guided by this insight, we propose \emph{ComE}, the first community embedding method so far as we know. 
We evaluate ComE on multiple real-world data sets, and show ComE outperforms the state-of-the-art baselines in both tasks of community prediction and node classification. Our code is available at \url{https://github.com/andompesta/nodeembedding-to-communityembedding}.
\end{abstract}

\section{Introduction} \label{sect:intro}

Traditionally, graph embedding focuses on individual \emph{nodes}, which aims to output a vector representation for each node in the graph, such that two nodes ``close'' on the graph have similar vector representations in a low-dimensional space. Such node embedding has been shown very successful in preserving the network structure, and significantly improving a wide range of applications, including node classification \cite{CaoLX15,PerozziAS2014}, node clustering \cite{DBLP:conf/aaai/TianGCCL14,DBLP:conf/ijcai/YangCHWWZ16}, link prediction \cite{GroverL16,OuCPZZ16}, graph visualization \cite{TangQWZYM15,DBLP:conf/kdd/WangC016} and more \cite{DBLP:conf/aaai/FangWZDZE16,DBLP:conf/icml/NiepertAK16}. 
\begin{figure}[t]
\centering
\includegraphics[width=\linewidth]{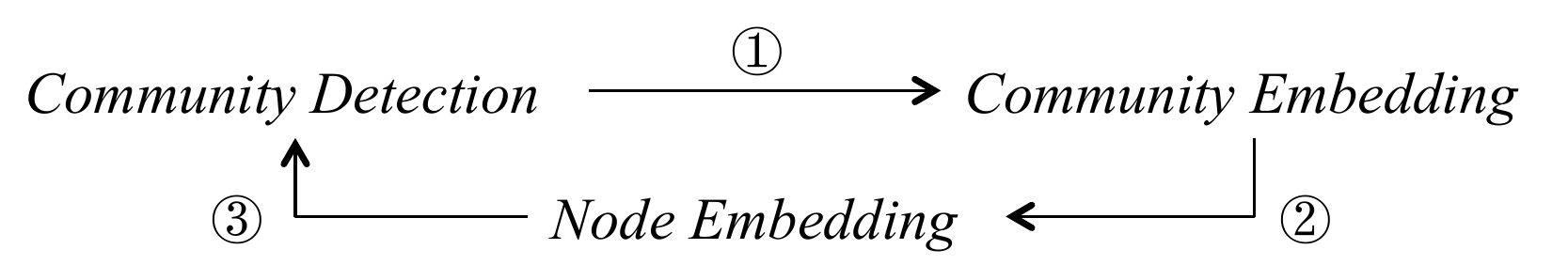}
\vspace{-6mm}
\caption{Close loop for community detection \& embedding.}
\vspace{-3mm}
\label{fig:loop}
\end{figure}

In this paper, we study a new setting for graph embedding, which focuses on embedding \emph{communities}. 
Generally, a ``community embedding'' is a representation for a community in a low-dimensional space. 
Since a community is a group of densely connected nodes, a community embedding cannot simply be a vector representation like a node embedding. Instead, a community embedding should be a set of \emph{random variables}, for a distribution characterizing how the community member nodes spread in the low-dimensional space. For instance, if we consider multivariate Gaussian as the distribution, then a community embedding consists of a mean and a covariance. Surprisingly, despite the success of node embedding, the concept of community embedding is still missing in the literature (more discussions in Sect. \ref{sec:related_work}).

\begin{figure*}[t]
  \centering
  \subfigure[Karate club graph]{
   \includegraphics[width=0.17\textwidth]{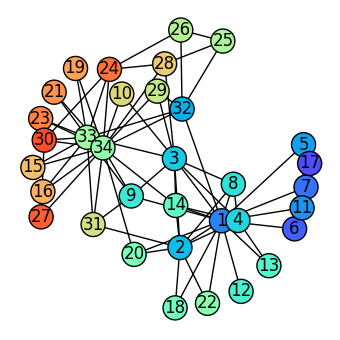}
   \label{figure.karate_graph}
   }
  \subfigure[DeepWalk]{
   \includegraphics[width=0.19\textwidth]{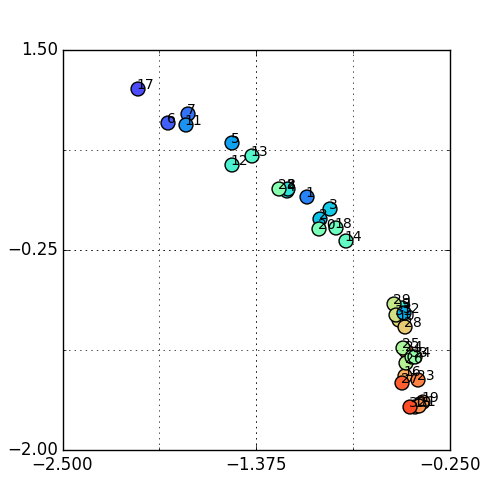}
   \label{figure.karate_deepwalk}
   }
  \subfigure[LINE]{
   \includegraphics[width=0.19\textwidth]{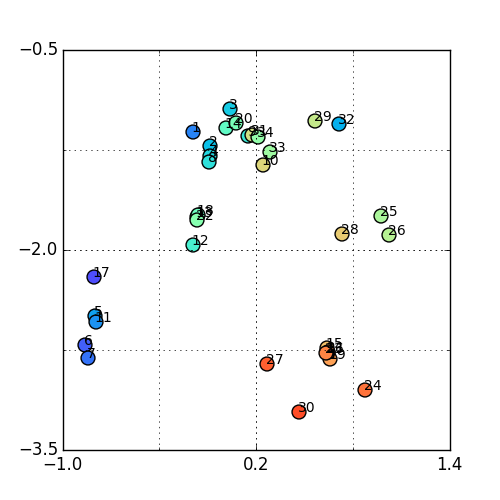}
   \label{figure.karate_line}
   }
  \subfigure[node2vec]{
   \includegraphics[width=0.19\textwidth]{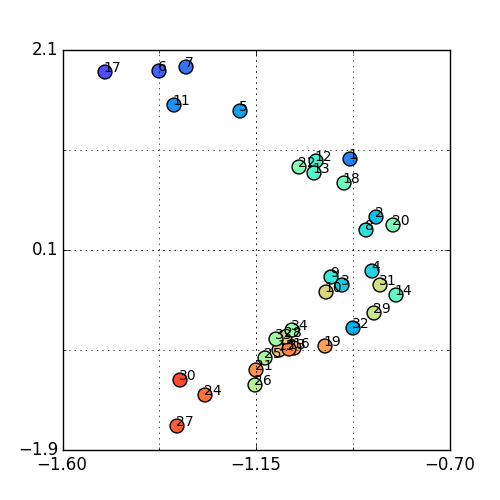}
   \label{figure.karate_node2vec}
   }
  \subfigure[ComE (ours)]{
   \includegraphics[width=0.19\textwidth]{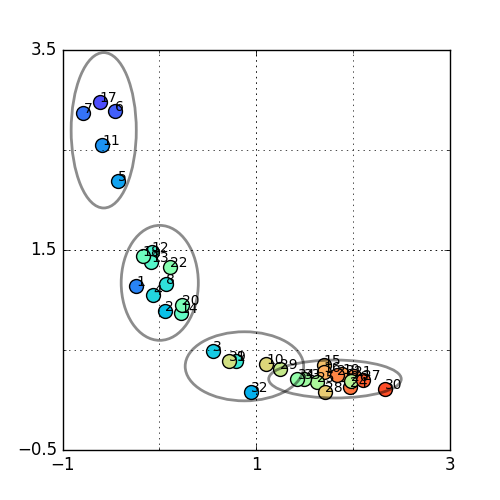}
   \label{figure.karate_our}
   }
  \caption{Graph embedding of the Zachary's karate club graph in a 2D space. 
In (b)--(e), we input the same data to different methods for graph embedding. For each node, we sample 10 paths, each of which has a length of 80. We set context window size as 5 and negative sampling size as 5. In LINE, we use mean pooling of the first-order proximity embedding and the second-order proximity embedding for each node. In node2vec, we set $p=0.25$, $q=0.25$. In ComE, we set $\alpha = 1$, $\beta = 1$. In (e), we also plot the community embeddings as four eclipses, characterizing the distribution of four detected communities.} 
\label{figure.karate}
\end{figure*}

Community embedding is useful. As a representation of community, it naturally supports community-level applications, such as visualizing communities to help generate insights for graph analysis, finding similar communities to assist community recommendation and so on. Besides, it also provides an exciting opportunity to improve community detection. Specifically, to find a useful community embedding, we first need to have a good community. Since recent graph embedding has shown to be an effective way for graph analytics, we wish to leverage node embedding to find good communities. Most recent node embedding methods such as DeepWalk \cite{PerozziAS2014}, LINE \cite{TangQWZYM15} and node2vec \cite{GroverL16} focus on preserving \emph{first-order} and/or \emph{second-order} proximity, which can ensure two nodes directly linked or sharing ``context'' to have similar embeddings. But they are not aware of community structures, as shown in Fig.~\ref{figure.karate_deepwalk}--\ref{figure.karate_node2vec}, where we visualize the node embedding for these methods on Zachary's karate club graph \cite{Zachary77}. 
Thus a simple pipeline approach to first detect communities and then aggregate their nodes' embeddings as community embedding (\ie, \ding{172} in Fig.~\ref{fig:loop}) is limited due to suboptimal detection. Community embedding can avoid such a limitation-- suppose we already have some community embedding, then we can come back to optimize the node embedding (\ie, \ding{173} in Fig.~\ref{fig:loop}), by ``squashing'' the nodes to scatter closely near their community centers. The resulting node embedding thus becomes community-aware and hopefully improves the community detection (\ie, \ding{174} in Fig.~\ref{fig:loop}). In Fig.~\ref{figure.karate_our}, we show a promising result to have community-aware node embedding and better communities, with the help of community embedding. 

To close the loop between community detection and community embedding is non-trivial. A shown in Fig.~\ref{fig:loop} \ding{173}, we try to use community embedding to guide node embedding, but it is not clear how to do it. There is work on using node embedding to improve community detection \cite{CaoLX15,DBLP:conf/nips/KozdobaM15,DBLP:conf/aaai/TianGCCL14}. But there is very little work on the other way around using explicit community information to enhance node embedding \cite{DBLP:conf/ijcai/YangCHWWZ16}, and it needs extra supervision of must-links to enforce the community in node embedding. 

Our insight for addressing the community embedding feedback is to see community embedding as providing a novel \emph{community-aware higher-order proximity}, such that two nodes in the same community can be close in the low-dimensional space. In contrast with first-order proximity and second-order proximity, community embedding does not require two nodes to be directly linked or share many ``contexts'' for being close. There is little work on higher-order proximity \cite{CaoLX15,OuCPZZ16}, and they all rely on embedding a higher-order adjacency matrix, which can easily have a quadratic number of non-zero entries. As a result, their computation can be expensive. Besides, their higher-order proximity does not explicitly model community. Another advantage of seeing community embedding as a higher-order proximity is that, we can combine it with the first-order and the second-order proximity for node embedding through the well-established neural network framework. 

Guided by the above insight, we propose \emph{ComE}, the first \underline{Com}munity \underline{E}mbedding model for graph analytics. To represent a community, we are inspired by the Gaussian Mixture Model \cite{Bishop06} to see each community as a Gaussian component. Thus we formulate a community embedding as a tuple of a mean vector indicating the center of a community and a covariance matrix indicating the spread of its members in a low-dimensional space. 
To realize the closed loop in Fig.~\ref{fig:loop}, we develop an iterative inference algorithm for \emph{ComE}. 
On the one hand, given node embeddings, 
we can detect and embed the communities, such that a good community assignment and a good community embedding should explain the node embedding well. 
On the other hand, given community assignment and community embedding, we can optimize the node embedding by enforcing all different orders of proximity. 
As shown in Sect. \ref{sect:infer}, our inference algorithm's complexity is linear to the graph size. 

We summarize our contributions as follows.
\vspace{-1mm}
{\flushleft $\bullet$ To the best of our knowledge, we are the first to introduce the concept of community embedding to graph analytics.}
\vspace{-1mm}
{\flushleft $\bullet$ We identify that communty embedding can improve community detection, thus closing the loop between them.}
\vspace{-1mm}
{\flushleft $\bullet$ We contribute a novel \emph{ComE} model to close the loop and a scalable inference algorithm to detect communities and infer their embeddings at the same time. }
\vspace{-1mm}
{\flushleft $\bullet$ We evaluate \emph{ComE} on three real-world data sets. It improves the state-of-the-art baselines by at least 2.5\%--7.8\% (NMI) and 1.1\%--2.8\% (conductance) in community detection, 9.52\%--43.5\% (macro-F1) and 6.9\%--19.4\% (micro-F1) in node classification.}

\section{Related Work} \label{sec:related_work}

There has been an increasing amount of graph data, ranging from social networks such as Twitter to various information networks such as Wikipedia. An important question in graph analytics is how to represent a graph. Graph embedding is a popular graph representation framework, which aims to project a graph into a low-dimensional space for further applications \cite{TenenbaumSL00,PerozziAS2014}. 

In terms of the target to embed, most graph embedding methods focus on nodes. 
For example, earlier methods, such as MDS \cite{coxc00}, LLE \cite{RoweisS00}, IsoMap \cite{TenenbaumSL00} and Laplacian eigenmap \cite{DBLP:conf/nips/BelkinN01}, typically aim to solve the leading eigenvectors of graph affinity matrices as node embedding. 
Recent methods typically rely on neural networks to learn the representation for each node, with either shallow architectures \cite{DBLP:conf/aaai/XieLJLS16,TangQWZYM15,GroverL16} or deep architectures \cite{DBLP:conf/icml/NiepertAK16,DBLP:conf/kdd/WangC016,DBLP:conf/kdd/ChangHTQAH15}.
Other than node embedding, there is some attempt to learn edge embedding in a knowledge base \cite{DBLP:conf/emnlp/LuoWWG15} or proximity embedding between two possibly distant nodes in a general heterogeneous graph \cite{LiuZZZCWY17}. But there is no community embedding so far as we know.

In terms of the information to preserve, most graph embedding methods try to preserve first-order proximity and/or second-order proximity \cite{PerozziAS2014,GroverL16,TangQWZYM15,DBLP:conf/kdd/WangC016}. Some recent attempts consider higher-order proximity, by factorizing a higher-order node-node proximity matrix by PageRank or Katz index \cite{CaoLX15,OuCPZZ16}. Hence their higher-order proximity is based on the graph reachability via random walk, where the notion of community is missing. In contrast, our community embedding tries to preserve all the first-, second- and community-aware higher-order proximity. 

In terms of interaction between node and community, there are a few graph embedding models that use node embedding to assist community detection \cite{DBLP:conf/aaai/TianGCCL14,DBLP:conf/nips/KozdobaM15}, but they do not have the notion of community in their node embedding. There is little work that allows community feedback to guide the node embedding \cite{DBLP:conf/ijcai/YangCHWWZ16}, but it lacks the concept of community embedding and its community feedback requires extra supervision on must-links. In contrast, we optimize node embedding, community embedding and community detection in a closed loop, and let them reinforce each other.

\section{Problem Formulation}

As \emph{input}, we are given a graph $G = (V,E)$, where $V$ is the node set and $E$ is the edge set. Traditional graph embedding aims to learn a node embedding for each $v_i \in V$ as $\bphi_i \in \mathbb{R}^d$. In this paper, we introduce the concept of community embedding. Suppose there are $K$ communities on the graph $G$. For each node $v_i$, we denote its community assignment as $z_i \in \{1, ..., K \}$. Motivated by Gaussian Mixture Model (GMM), we represent each community as a Gaussian component, which is characterized by a mean vector indicating the community center and a covariance matrix indicating its member nodes' spread. 
Formally, we define:

\begin{definition} \label{def.community_embedding}
A community $k$'s embedding (where $k=1,2,..., K$) in a $d$-dimensional space is a set of random variables $(\bpsi_k, \Sigma_k)$, where $\bpsi_k \in \mathbb{R}^d$ is the mean and $\Sigma_k \in \mathbb{R}^{d \times d}$ is the covariance for a multivariate Gaussian distribution. 
\end{definition}

As \emph{output}, we aim to learn both the community embedding $(\bpsi_k, \Sigma_k)$ for each community $k \in \{1, ..., K\}$ and the node embedding $\bphi_i$ for each node $v_i \in V$ on the graph $G$. 

Next, we model the closed loop in Fig.~\ref{fig:loop}. 

\subsection{Community Detection and Embedding} \label{sect:community_embedding}

Given node embedding, one straightforward way to detect communities and learn their embedding is to take a pipeline approach. For example, as shown in Fig.~\ref{fig:loop}, one can first run K-means to detect communities, and then fit a Gaussian mixture for each community. However, such a pipeline approach lacks a unified objective function, thus hard to optimize later with node embedding. 
Alternatively, we can do community detection and embedding together in one single objective function based on Gaussian Mixture Model. That is, we consider each node $v_i$'s embedding $\bphi_i$ as generated by a multivariate Gaussian distribution from a community $z_i = k$. Then, for all the nodes in $V$, we have the likelihood as
\begin{equation} \label{eq.community}
  \textstyle \prod\nolimits_{i=1}^{|V|} \sum\nolimits_{k=1}^K p(z_i = k) p(v_i|z_i = k; \bphi_i, \bpsi_k, \Sigma_k),
\end{equation}
where $p(z_i = k)$ is the probability of node $v_i$ belonging to community $k$. For notation simplicity, we denote $p(z_i = k)$ as $\pi_{ik}$; thus we have $\pi_{ik} \in [0,1]$ and $\sum_{k=1}^K \pi_{ik} = 1$. 
In community detection, these $\pi_{ik}$'s indicate the \emph{mixed community membership} for each node $v_i$, and they are unknown. 
Besides, $p(v_i|z_i = k; \bphi_i, \bpsi_k, \Sigma_k)$ is a multivariate Gaussian distribution defined as follows
\begin{equation}
  p(v_i|z_i = k; \bphi_i, \bpsi_k, \Sigma_k) = \mathcal{N}(\bphi_i | \bpsi_k, \Sigma_k).
\end{equation}
In community embedding, the $(\bpsi_k, \Sigma_k)$'s are unknown. 
By optimizing Eq.~\ref{eq.community} \wrt $\pi_{ik}$'s and $(\bpsi_k, \Sigma_k)$'s, we achieve community detection and embedding at the same time. 

\subsection{Node Embedding} \label{sect:node_embedding}

Traditionally, node embedding focuses on preserving first- or second-order proximity. 
For example, to preserve first-order proximity, LINE \cite{TangQWZYM15} enforces two neighboring nodes to have similar embedding by minimizing 
\vspace{-2mm}
\begin{equation} \label{eq:o1}
\vspace{-1mm}
 O_1 =\textstyle -\sum\nolimits_{(v_i,v_j) \in E} \log \sigma(\bphi_j^T \bphi_i),
\end{equation}
where $\sigma(x) = 1/(1 + \exp(-x))$ is a sigmoid function. 

To preserve second-order proximity, LINE and DeepWalk \cite{PerozziAS2014} both enforce two nodes sharing many ``contexts'' (i.e., neighbors within $\zeta$ hops) to have similar embedding. In this case, each node has two roles: a node for itself and a context for some other nodes. To differentiate such roles, LINE introduces an extra context embedding for each node $v_j$ as $\bphi'_j \in \mathbb{R}^d$. Denote $C_i$ as the set of contexts for node $v_i$. Then LINE adopts \emph{negative sampling} \cite{DBLP:conf/nips/MikolovSCCD13} to define a function for measuring how well $v_i$ generates each of its contexts $v_j \in C_i$ as
\vspace{-2mm}
\begin{equation} \label{eq.delta}
\vspace{-1mm}
  \Delta_{ij} \textstyle = \log \sigma({\bphi'}_j^T \bphi_i) + \sum\limits_{t=1}^{m} \mathbb{E}_{v_l \sim P_n(v_l)}[\log \sigma(-{\bphi'}_l^T \bphi_i)],
\end{equation}
where $v_l \sim P_n(v_l)$ denotes sampling a node $v_l \in V$ as a ``negative context'' of $v_i$ according to a probability $P_n(v_l)$. 
We set $P_n(v_l) \propto r_l^{3/4}$ as proposed in \cite{DBLP:conf/nips/MikolovSCCD13}, where $r_l$ is $v_l$'s degree. In total, there are $m$ negative contexts. Generally, maximizing Eq.~\ref{eq.delta} enforces node $v_i$'s embedding $\bphi_i$ to best generate its positive contexts $\bphi_j'$'s, but not its negative contexts $\bphi_l'$'s. Then we can minimize the following objective function to preserve the second-order proximity:
\vspace{-2mm}
\begin{equation} \label{eq:o2}
\vspace{-1mm}
  O_2 = \textstyle - \alpha \sum\nolimits_{v_i \in V} \sum\nolimits_{v_j \in C_i} \Delta_{ij},
\end{equation}
where $\alpha > 0$ is a trade-off parameter.

\subsection{Closing the Loop}

In order to close the loop in Fig.~\ref{fig:loop}, we need to enable the feedback from community detection and community embedding to node embedding. 
Suppose we have identified the mixed community membership $\pi_{ik}$'s and the community embedding $(\bpsi_k, \Sigma_k)$'s in Sect.~\ref{sect:community_embedding}. Then we can re-use Eq.~\ref{eq.community} to enable such feedback, by seeing the node embedding $\bphi_i$'s as \emph{unknown}. Effectively, optimizing Eq.~\ref{eq.community} \wrt $\bphi_i$'s enforces the nodes $\bphi_i$'s within the same community to get closer to the corresponding community center $\bpsi_k$. That is, two nodes sharing a community are likely to have similar embedding. Compared with the first- and second-order proximity, this design introduces a new \emph{community-aware higher-order proximity} to node embedding, which is useful for community detection and embedding later. 
For example, in Fig.~\ref{figure.karate_graph}, node 3 and node 10 are directly linked, but node 3 is closer to the blue-green community, whereas node 10 is closer to the red-yellow community. Therefore, by only preserving first-order proximity, we may not tell their community membership's difference well. 
For another example, node 9 and node 10 share a number of one-hop and two-hop neighbors, but compared with node 10, node 9 is closer to the blue-green community. 
Therefore, by only preserving second-order proximity, we may not tell their community membership's difference well, as shown in Fig.~\ref{figure.karate_deepwalk} and Fig.~\ref{figure.karate_node2vec}. 

Based on the closed loop, we will optimize community detection, community embedding and node embedding together. 
We have three types of proximity to consider for node embedding, including first-, second- and higher-order proximity. 
In general, there are two approaches to combine different types of proximity for node embedding: 
1) ``concatenation'', \eg, LINE first separately optimizes $O_1$ and $O_2$, then it concatenates the two resulting  embedding for each node into a long vector as the final output; 
2) ``unification'', \eg, SDNE \cite{DBLP:conf/kdd/WangC016} learns a single node embedding for each node to preserve both first- and second-order proximity at the same time. 
In this paper, to encourage the node embedding to unify multiple types of proximity, we adopt the unification approach, and leave the other approach as future work. 
Consequently, based on Eq.~\ref{eq.community}, we first define the objective function for community detection and embedding, as well as enforcing the higher-order proximity for node embedding as:
\vspace{-2mm}
\begin{equation} \label{eq:o3}
\vspace{-1mm}
  O_3 =\textstyle - \frac{\beta}{K} \sum\nolimits_{i=1}^{|V|} \log \sum\nolimits_{k=1}^K \pi_{ik} \mathcal{N}(\bphi_i | \bpsi_k, \Sigma_k),
\end{equation}
where $\beta \geq 0$ is a trade-off parameter. 
Denote $\Phi = \{ \bphi_i \}$, $\Phi' = \{ \bphi'_i \}$, $\Pi = \{\pi_{ik} \}$, $\Psi = \{ \bpsi'_k \}$ and $\Sigma = \{\Sigma_k \}$ for $i =1,...,N, k=1,...,K$. 
Then, we also unify the first- and second-order proximity for node embedding, thus reaching the ultimate objective function for \emph{ComE} as 
\begin{equation} \label{eq.overall_obj}
   \mathcal{L}(\Phi, \Phi', \Pi, \Psi, \Sigma) =\textstyle O_1(\Phi) + O_2(\Phi, \Phi') + O_3(\Phi, \Pi, \Psi, \Sigma).
\end{equation}
Our final optimization problem becomes:
\begin{equation} \label{eq.overall_opt}
  (\Phi^*, \Phi'^*, \Pi^*, \Psi^*, \Sigma^*) \leftarrow \underset{\forall k, diag(\Sigma_k) > \mathbf{0}}{\arg\min} \mathcal{L}(\Phi, \Phi', \Pi, \Psi, \Sigma),
\end{equation}
where $diag(\Sigma_k)$ returns the diagonal entries of $\Sigma_k$. We particularly introduce a constraint of $diag(\Sigma_k) > \mathbf{0}$ for each $k \in \{1, ..., K\}$ to avoid the singularity issue of optimizing $\mathcal{L}$. Similar to Gaussian Mixture model (\cite{Bishop06}, there exists degenerated solutions for optimizing $\mathcal{L}$ without any constraint. That is, when a Gaussian component collapses to a single point, the $diag(\Sigma_k)$ becomes zero, which makes $O_3$ become negative infinity.

\section{Inference} \label{sect:infer}
We decompose the optimization of Eq.~\ref{eq.overall_opt} into two parts, and take an iterative approach to solve it. 
Specifically, we consider iteratively optimizing $(\Pi, \Psi, \Sigma)$ with a constrained minimization given $(\Phi, \Phi')$, and optimizing $(\Phi, \Phi')$ with an unconstrained minimization given $(\Pi, \Psi, \Sigma)$. Empirically, this iterative optimization algorithm converges quickly with a reasonable initialization; \eg, we initialize $(\Phi, \Phi')$ by DeepWalk results in our experiments. We report the convergence in Sect.~\ref{sect:exp_model_study}. Next we detail this iterative optimization. 

\stitle{Fix $(\Phi, \Phi')$, optimize $(\Pi, \Psi, \Sigma)$}. In this case, Eq.~\ref{eq.overall_obj} is simplified as the negative log-likelihood of a GMM. According to \cite{Bishop06}, we can optimize $(\Pi, \Psi, \Sigma)$ by EM, and obtain a closed-form solution as
\begin{align}
  \pi_{ik} &= \textstyle \frac{N_k}{|V|}, \label{eq:pi_ik} \\
  \bpsi_k &= \textstyle \frac{1}{N_k} \sum\nolimits_{i=1}^{|V|} \gamma_{ik} \bphi_i, \label{eq:psi_k} \\ 
  \Sigma_k &= \textstyle \frac{1}{N_k} \sum\nolimits_{i=1}^{|V|} \gamma_{ik} (\bphi_i - \bpsi_k) (\bphi_i - \bpsi_k)^T, \label{eq:Sigma_k}
\end{align}
where $\gamma_{ik} = \frac{\pi_{ik} \mathcal{N}(\bphi_i|\bpsi_k,\Sigma_k)}{\sum_{k'=1}^K \pi_{ik'} \mathcal{N}(\bphi_i|\bpsi_{k'},\Sigma_{k'})}$ and $N_k = \sum\nolimits_{i=1}^{|V|} \gamma_{ik}$.

\stitle{Fix $(\Pi, \Psi, \Sigma)$, optimize $(\Phi, \Phi')$}. In this case, Eq.~\ref{eq.overall_obj} is simplified as optimizing the node embedding with three types of proximity. Due to the summation within the logarithm term of $O_3$, it is inconvenient to compute the gradient of $\bphi_i$. Thus we try to minimize an upper bound of $\mathcal{L}(\Phi, \Psi, \Sigma)$ instead:
\begin{equation}
  O_3' = \textstyle - \frac{\beta}{K} \sum\nolimits_{i=1}^{|V|} \sum\nolimits_{k=1}^K \pi_{ik} \log \mathcal{N}(\bphi_i | \bpsi_k, \Sigma_k).
\end{equation}
It is easy to prove that
  $O_3'(\Phi; \Pi, \Psi, \Sigma) \geq O_3(\Phi; \Pi, \Psi, \Sigma)$,  
due to log-concavity $\sum\nolimits_{i=1}^{|V|} \log \sum\nolimits_{k=1}^K \pi_{ik} \mathcal{N}(\bphi_i | \bpsi_k, \Sigma_k) \geq \sum\nolimits_{i=1}^{|V|} \sum\nolimits_{k=1}^K \log \pi_{ik} \mathcal{N}(\bphi_i | \bpsi_k, \Sigma_k)$.
As a result, we define  
\[
  \mathcal{L}'(\Phi, \Phi') =  O_1(\Phi) + O_2(\Phi, \Phi') + O_3'(\Phi; \Pi, \Psi, \Sigma),
\] 
and thus $ \mathcal{L}'(\Phi, \Phi') \geq \mathcal{L}(\Phi, \Phi')$. 
We optimize $\mathcal{L}'(\Phi, \Phi') $ by \emph{stochastic gradient descent} (SGD). For each $v_i \in V$, we have
\begin{align}
   \textstyle \frac{\partial O_1}{\partial \bphi_i}  &\textstyle= -\sum\nolimits_{(i,j) \in E} \sigma(-\bphi_j^T \bphi_i) \bphi_j,    \label{eq:O1_phi_i} \\
   \textstyle \frac{\partial O_2}{\partial \bphi_i}  &\textstyle= - \alpha  \sum\nolimits_{v_j \in C_i} \bigg[ \sigma(-{\bphi_j'}^T \bphi_i) {\bphi_j'} \nonumber \\
   &~~~~~~~~~~~~~~\textstyle+ \sum\nolimits_{t=1}^{m} \mathbb{E}_{v_l \sim P_n(v_l)}[ \sigma( {\bphi_l'}^T \bphi_i) (- \bphi_l') ] \bigg],  \label{eq:O2_phi_i} \\
   \textstyle \frac{\partial O_3'}{\partial \bphi_i} &\textstyle= \frac{\beta}{K} \sum\nolimits_{k=1}^K \pi_{ik} \Sigma_k^{-1} (\bphi_i - \bpsi_k). \label{eq:O3'_phi_i}
\end{align}
We also compute the gradient for context embedding as
\begin{align}
   &\textstyle \frac{\partial O_2}{\partial \bphi_j'} 
   =\textstyle - \alpha \sum\nolimits_{v_i \in V} \bigg[ \delta(v_j \in C_i) \sigma(-{\bphi_j'}^T \bphi_i) {\bphi_i} \nonumber \\
   &\textstyle~~~+ \sum\nolimits_{t=1}^{m} \mathbb{E}_{v_l \sim P_n(v_l)}[ \delta(v_l = v_j) \sigma({\bphi_l'}^T \bphi_i) (- \bphi_i)] \bigg]. \label{eq:phi_j'}
\end{align}

\vspace{-3mm}
\stitle{Algorithm and complexity}. 
We summarize ComE in Alg.~\ref{algorithm}.
In line 1, for each $v_i \in V$, we sample $\gamma$ paths starting from $v_i$ with length $\ell$ on $G$. 
In lines 4--5, we fix $(\Phi, \Phi')$ and optimize $(\Pi, \Psi, \Sigma)$ for community detection and embedding.
In lines 6--13, we fix $(\Pi, \Psi, \Sigma)$ and optimize $(\Phi, \Phi')$ for node embedding. 
Specifically, we update node embedding by first-order proximity (lines 6--7), second-order proximity (lines 8--11) and community-aware higher-order proximity (lines 12--13).

\begin{algorithm}[t]
\caption{ComE}
\label{algorithm}
\begin{algorithmic}[1]
    \REQUIRE graph $G = (V,E)$, \#(community) $K$, \#(paths per node) $\gamma$, walk length $\ell$, context size $\zeta$, embedding dimension $d$, negative context size $m$, parameters $(\alpha, \beta)$.
   \ENSURE node embedding $\Phi$, context embedding $\Phi'$, community assignment $\Pi$, community embedding $(\Psi, \Sigma)$.

   \STATE $\mathcal{P} \leftarrow$ SamplePath($G, \ell$);
    \STATE Initialize $\Phi$ and $\Phi'$;
    \FOR{$iter=1:T_1$}
   	 \FOR{$subiter=1:T_2$}
		\STATE Update $\pi_{ik}$, $\bpsi_k$ and $\Sigma_k$ by Eq.~\ref{eq:pi_ik}, Eq.~\ref{eq:psi_k} and Eq.~\ref{eq:Sigma_k};
 	   \ENDFOR
	\FORALL{edge $(i, j) \in E$}
		\STATE SGD on $\bphi_i$ and $\bphi_j$ by Eq.~\ref{eq:O1_phi_i}; 
	\ENDFOR 
	\FORALL{path $p \in \mathcal{P}$}
		\FORALL{$v_i$ in path $p$}
			\STATE SGD on $\bphi_i$ by Eq.~\ref{eq:O2_phi_i};
			\STATE SGD on its context $\bphi'_j$'s within $\zeta$ hops by Eq.~\ref{eq:phi_j'};
		\ENDFOR
	\ENDFOR
	\FORALL{node $v_i \in V$}
		\STATE SGD on $\bphi_i$ by Eq.~\ref{eq:O3'_phi_i};
	\ENDFOR
    \ENDFOR
\end{algorithmic}
\end{algorithm}

We analyze time complexity of Alg.~\ref{algorithm}. 
Path sampling in line 1 takes $O(|V| \gamma \ell)$. 
Parameter initialization in line 2 takes $O(|V| + K)$. 
Community detection and embedding in line 5 takes $O(|V|K)$. 
Node embedding \wrt first-order proximity in lines 6--7 takes $O(|E|)$.
Node embedding \wrt second-order proximity in lines 8--11 takes $O(|V| \gamma \ell)$.
Node embedding \wrt community-aware higher-order proximity in lines 12--13 takes $O(|V| K)$.  
In total, the complexity is $O(|V| \gamma \ell + |V| + K + T_1 \times (T_2 |V| K + |E| + |V| \gamma \ell + |V| K))$, which is linear to the graph size (i.e., $|V|$ and $|E|$). 

\section{Experiments}
We want to quantitatively evaluate community embedding. 
Firstly, as an important application of community embedding, we evaluate the resulting community detection (Sect.~\ref{sect:exp_community_detection}). 
Secondly, as the loop between community detection and embedding is closed through node embedding,  we also evaluate the resulting node embedding. In particular, following the previous work \cite{PerozziAS2014,TangQWZYM15}, we consider node classification (Sect.~\ref{sect:exp_node_classification}). 
Finally, we study our model's convergence and parameter sensitivity (Sect.~\ref{sect:exp_model_study}). 

\stitle{Data sets}. We use three public data sets\footnote{Available at http://socialcomputing.asu.edu/pages/datasets and https://snap.stanford.edu/node2vec/POS.mat.}, as listed in Table \ref{table.datasets}.
\begin{table}[h]
\centering
\small
\addtolength{\tabcolsep}{-3pt}
\caption{Data sets.} \label{table.datasets}
\begin{tabular}{c|c|c|c|c}
 \hline
  & \#(node) & \#(edge) & \#(node labels) & labels per node \\ \hline
 BlogCatalog & 10,312 & 333,983 & 39 & single-label \\ \hline
 Flickr & 80,513 & 5,899,882 & 195 & single-label \\ \hline
 Wikipedia & 4,777 & 184,812 & 40 & multi-label \\ \hline
\end{tabular}
\end{table}

\vspace{-2mm}
\stitle{Evaluation metrics}. In community detection, we use both \emph{conductance} \cite{KlosterG14} and \emph{normalized mutual information} (NMI) \cite{DBLP:conf/aaai/TianGCCL14}. 
In node classification, we use \emph{micro-F1} and \emph{macro-F1} \cite{PerozziAS2014}. 

\stitle{Baselines}. 
We design baselines to answer two questions: 
1) is it necessary to model community detection, community embedding and node embedding together? 
2) is it necessary to enable the feedback from community embedding to community detection through node embedding? 
To answer the first question, we introduce a simple pipeline approach. 

\vspace{-1mm}
{\flushleft $\bullet$ \emph{Pipeline}: it first detects communities by GMM based on the graph adjacency matrix. Then, it uses LINE to generate first- and second-order node embedding. Finally, it fits a multivariate Gaussian in each community based on its member nodes' embedding as the community embedding.}

To answer the second question, we apply the state-of-the-art methods to generate node embedding, based on which we further use GMM to detect and embed communities.
{\flushleft $\bullet$ \emph{DeepWalk} \cite{PerozziAS2014}: it models second-order proximity for node embedding.}
{\flushleft $\bullet$ \emph{LINE} \cite{TangQWZYM15}: it models both first- and second-order proximity for node embedding.}
{\flushleft $\bullet$ \emph{node2vec} \cite{GroverL16}: it extends DeepWalk by exploiting homophily and structural roles.}
{\flushleft $\bullet$ \emph{GraRep} \cite{CaoLX15}: it models random walk based higher-order proximity for node embedding.}
{\flushleft $\bullet$ \emph{SAE} \cite{DBLP:conf/aaai/TianGCCL14}: it uses sparse auto-encoder, which is shown to share a similar objective as spectral clustering, on the adjacency matrix for node embedding.}

We try to compare with all baselines on all data sets, using the codes released by their authors. However, we cannot produce results for \emph{node2vec} and \emph{GraRep} on Flickr due to unmanageable out-of-memory errors on a machine with 64GB memory. Thus we exclude them from comparison on Flickr. 

\stitle{Parameters and environment}. In \emph{DeepWalk}, \emph{node2vec} and ComE, we set $\gamma = \zeta = 10$, $\ell = 80$, $m = 5$. 
In \emph{node2vec}, we set $p= q=0.25$ according to the best results in \cite{GroverL16}. We set the embedding dimension $d = 128$ for all methods. We set $K$ as the number of unique labels in each data set. 
We run experiments on Linux machines with eight 3.50GHz Intel Xeon(R) CPUs and 16GB memory.

\subsection{Community Detection}  \label{sect:exp_community_detection}
\begin{figure}[t]
\centering
\includegraphics[width=1\linewidth]{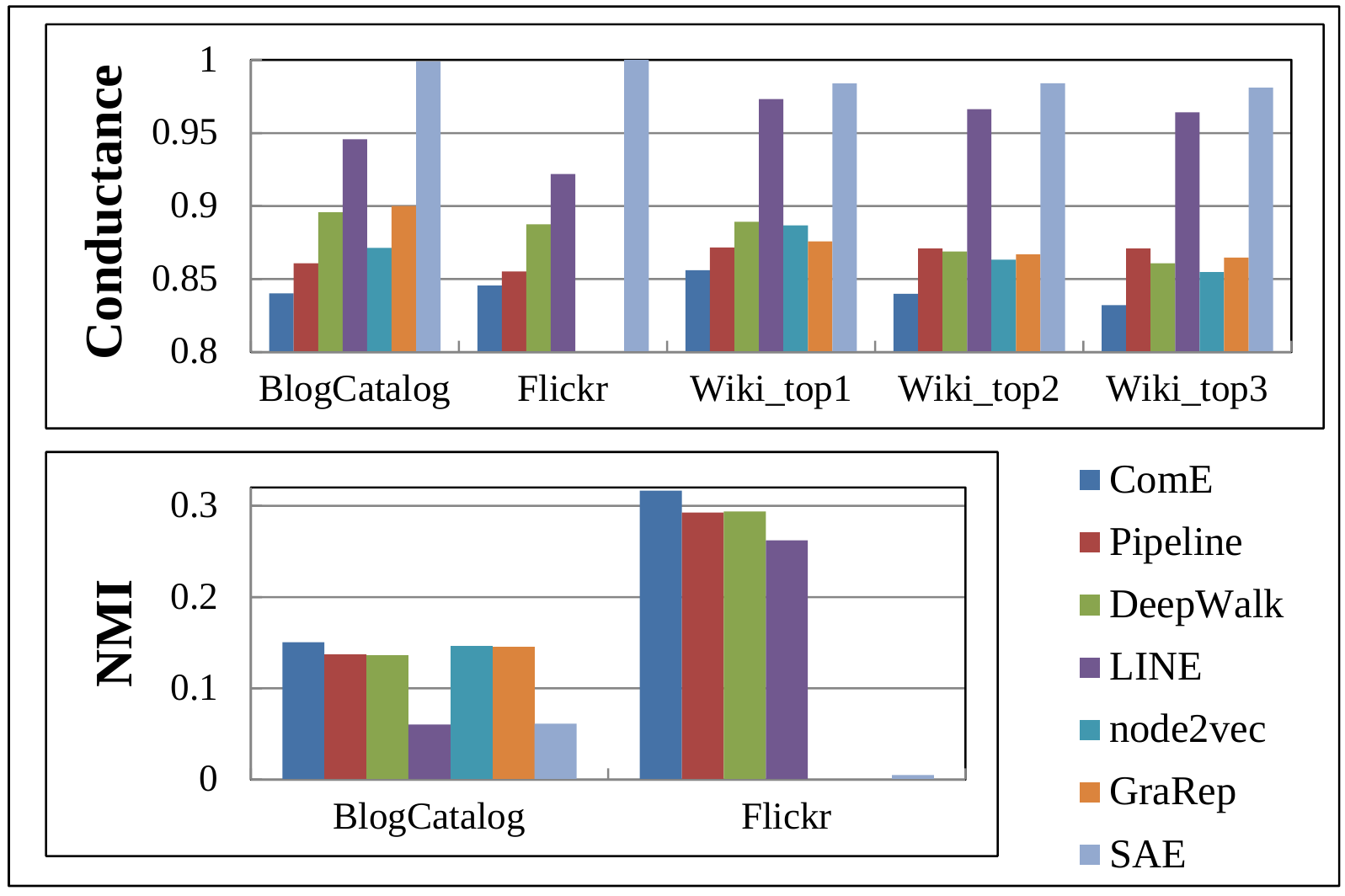}
\caption{Results on community prediction. The smaller conductance is, the better. The bigger NMI is, the better.}
\label{fig:clustering}
\end{figure}
\begin{figure}[t]
\centering
\includegraphics[width=1\linewidth]{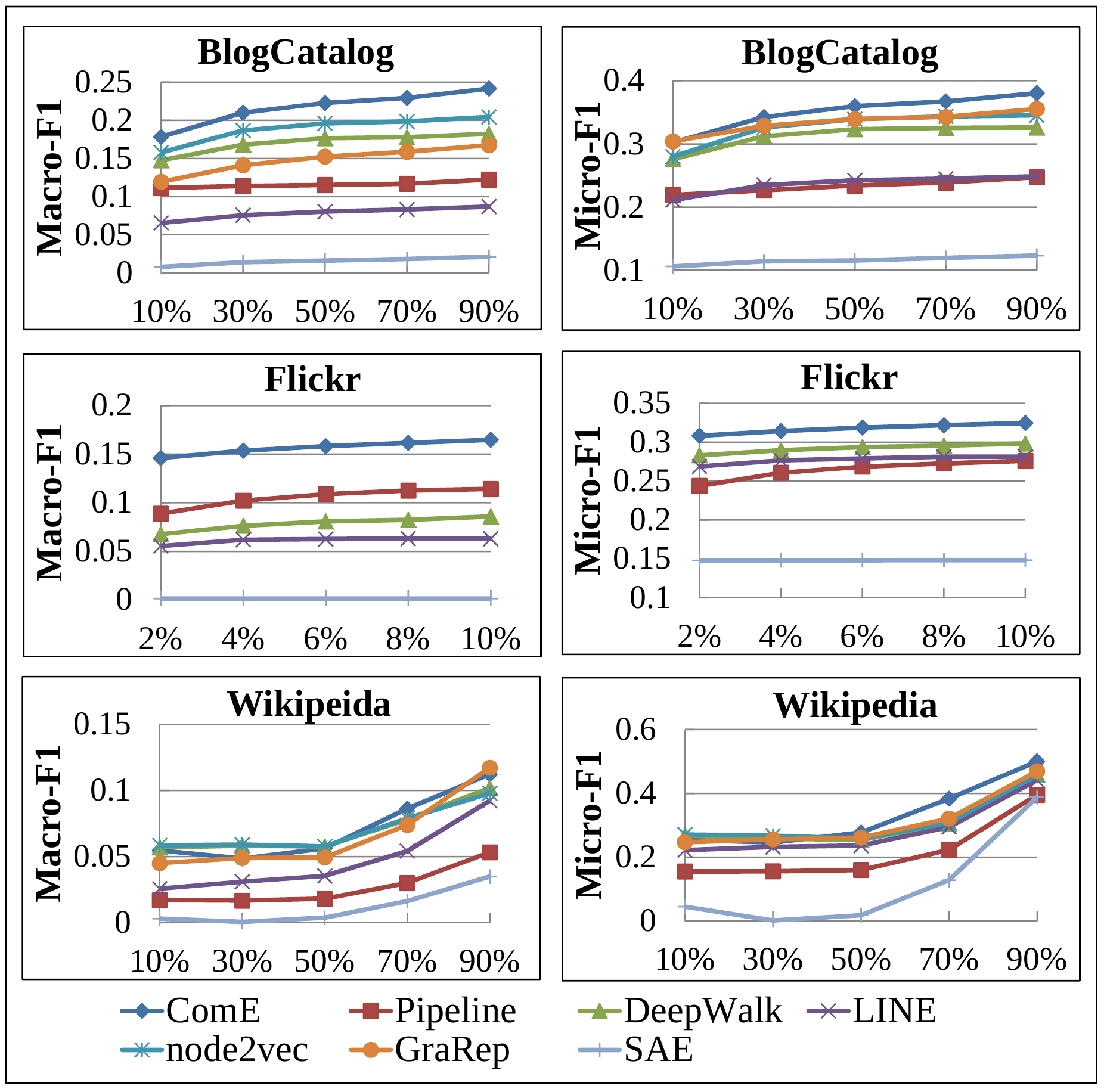}
\caption{Results on node classification. The bigger macro-F1 and micro-F1 are, the better.}
\label{fig:classification}
\end{figure}
\begin{figure}[t]
\centering
\includegraphics[width=1\linewidth]{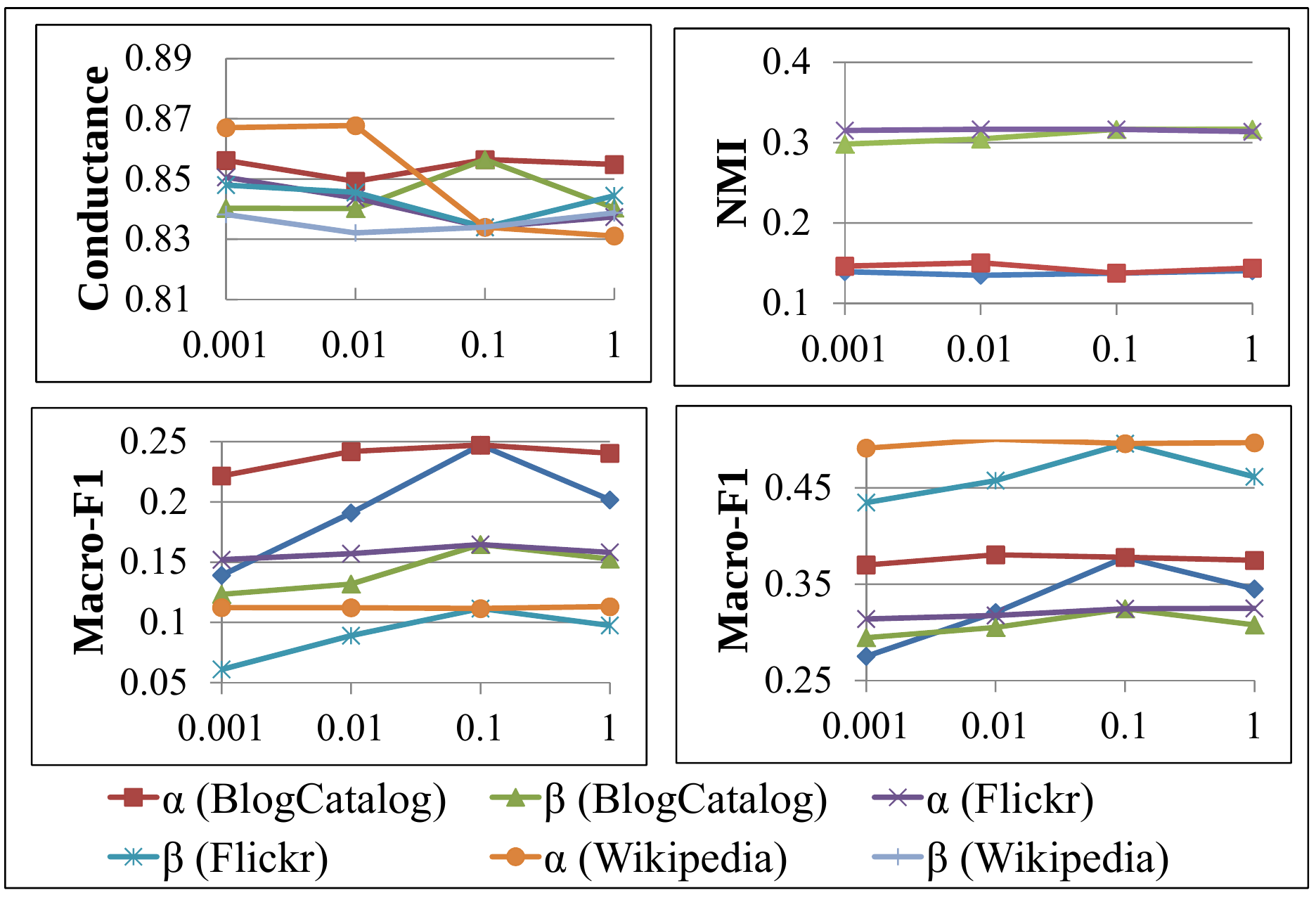}
\caption{Impact of the parameters $\alpha$ and $\beta$. By default, $\alpha=0.1$, $\beta=0.1$.}
\label{fig:para}
\end{figure}

\begin{figure}[t]
\centering
\includegraphics[width=1\linewidth]{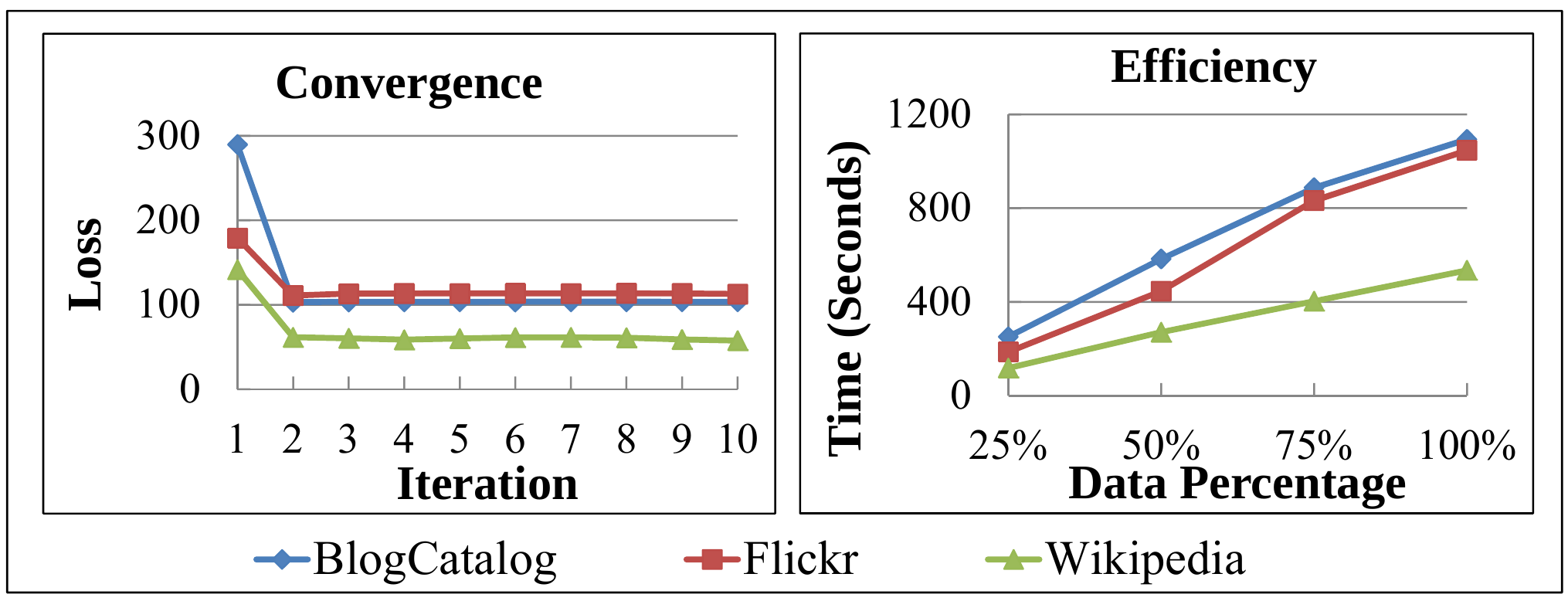}
\caption{Model's Convergence and Efficiency}
\label{fig:cov}
\end{figure}
In community prediction, our goal is to predict the most likely community assignment for each node. Since our data sets are labeled, we set the number of communities $K$ as the number of distinct labels in the data set. As an unsupervised task, we use the whole graph for learning embeddings and then predicting communities for each node. Note that in Wikipedia, one node has multiple labels, thus we predict the top $N$($N=1,2,3$) communities for each node, the results are denoted as Wiki$\_$top$N$($N=1,2,3$) in Fig.~\ref{fig:clustering}.

We compare ComE with the baselines for community prediction. We set $\alpha = 0.1, \beta = 0.01$ in ComE for all three data sets. As the baselines do not consider community in their embeddings, for fair comparison, we apply GMM over all the methods' node embedding outputs for community prediction. 
As shown in Fig.~\ref{fig:clustering}, ComE is consistently better than the baselines in terms of both conductance and NMI. Specifically, ComE improves the best baseline in all data sets by relative 1.1\%--2.8\% (conductance) and 2.5\%--7.8\% (NMI). This improvement suggests that, for community prediction, modeling community together with node embedding is better than doing them separately. Note that we do not calculate NMI for Wikipeida as it is multilabel dataset.

\subsection{Node Classification} \label{sect:exp_node_classification}

In node classification, our goal is to classify each node into one (or more) labels. 
We follow \cite{PerozziAS2014} to first train the embeddings on the whole graph, then randomly split 10\% (BlogCatalog and Wikipedia) and 90\% (Flickr) of nodes as test data, respectively. We use the remaining nodes to train a classifier by LibSVM ($c=1$ for all methods) \cite{ChangL11}. We repeat 10 splits and report the average results. 

We compare ComE with the baselines for node classification. We set $\alpha=0.1$, $\beta = 0.01$ for BlogCatalog and Wikipedia, and $\alpha = 0.1$, $\beta = 0.1$ for Flickr. We vary the number of training data to build the classifiers for each method's node embeddings. As shown in Fig.~\ref{fig:classification}, ComE is generally better than the baselines in terms of both macro-F1 and micro-F1. Specifically, ComE improves the best baselines by relatively 9.52\%--43.5\% (macro-F1) and 6.9\%--19.4\% (micro-F1), when using 70\% (BlogCatalog and Wikipedia) and 8\% (Flickr) of nodes for training. Our student t-tests show that all the above relative improvements are significant over the 10 data splits, with one-tailed $p$-values always less than 0.01.  
It is interesting to see ComE improves the baselines on node classification, since community embedding is after all unsupervised and it does not directly optimize the classification loss. This suggests that the higher-order proximity from community embedding does contribute to a better node embedding.

\subsection{Model Study} \label{sect:exp_model_study}

We test different values for the model parameters $\alpha$ and $\beta$. As shown in Fig.~\ref{fig:para}, generally $\alpha = 0.1$ gives the best results for community prediction and node classification. This suggests keeping an appropriate trade off for the second-order proximity in the objective function is necessary. $\beta = 0.1$ is generally the best for both tasks. In general, when $\alpha$ and $\beta$ are within the range of $[0.001, 1]$, the model performance is quite robust. We further validate the convergence and efficiency of ComE. As shown in Fig \ref{fig:cov}, the loss of ComE (normalized by $|V|$) converges quickly within a few iterations and the processing time of ComE is linear to the graph size (i.e., $|V|$ and $|E|$).

\section{Conclusion}
In this paper, we study the problem of embedding communities on graph. The problem is new because most graph embedding methods focus on individual nodes, instead of a group of nodes. 
We observe that community embedding and node embedding reinforce each other. 
On one hand, a good community embedding helps to get a good node embedding, because it preserves the community structure during embedding. 
On the other hand, a good node embedding also helps to get a good community embedding, as clustering is then done over the nodes with good representations. 
We jointly optimize node embedding and community embedding. We evaluate our method on the real-world data sets, and show that it outperforms the state-of-the-art baselines by at least 2.5\%--7.8\% (NMI) and 1.1\%--2.8\% (conductance) in community prediction, 9.52\%--43.5\% (macro-F1) and 6.9\%--19.4\% (micro-F1) in node classification.

\newpage

\bibliographystyle{named}
\bibliography{community_embedding} 

\end{document}